\documentclass[]{spie}  %>>> use this for US letter paper

\usepackage[english]{babel}
\usepackage{amsmath,amsfonts}
\usepackage{graphicx}
\usepackage[colorlinks=true, allcolors=blue]{hyperref}

\providecommand{\arcsec}{\ensuremath{^{\prime\prime}}}
\providecommand{\arcmin}{\ensuremath{^{\prime}}}

\title{The Constant Evolution of the Neil Gehrels Swift Observatory: What Two Decades of Operational Innovation Have Achieved for Time-Domain and Multi-Messenger Astrophysics}

\author{Jamie A. Kennea*\\
Astrophysics \& Space Center, Schmidt Sciences, New York, NY 10011, USA}

\authorinfo{*E-mail: jkennea@schmidtsciences.org}

\begin{document}
\maketitle

\begin{abstract}
The Neil Gehrels Swift Observatory has, for more than twenty years, served as NASA's premier rapid-response, multi-wavelength facility for the study of gamma-ray bursts (GRBs) and the transient sky. While Swift's hardware has remained essentially unchanged since its 2004 launch, its operational capabilities have evolved continuously, driven by a philosophy of ``always be developing.'' Hosted entirely at The Pennsylvania State University, with co-located flight and science operations, Swift has repeatedly reinvented how a NASA mission can be commanded, transforming itself from a GRB-chasing autonomous robotic telescope into a flexible, automated platform for time-domain and multi-messenger astrophysics (TDAMM). We review the key operational innovations that have enabled this evolution: the development of automated target-of-opportunity (TOO) uploads in response to the early loss of active X-ray Telescope cooling; onboard tiling of large error regions; the ManyPoint flight-software capability for executing hundreds of pointings; the Gamma-ray Urgent Archiver for Novel Opportunities (GUANO); and the ``Urgency 0'' continuous-commanding mode that has reduced TOO response latencies from hours to seconds. We illustrate the scientific impact of these capabilities through Swift's gravitational-wave and neutrino follow-up campaigns, including GW170817/AT2017gfo and IceCube-170922A, and through real-time ``early-warning'' slewing toward predicted compact-binary mergers. Finally, we discuss the heritage Swift is creating for future TDAMM missions, and the ongoing efforts to extend Swift's life, both through attitude-based scheduling that minimizes the spacecraft's atmospheric drag cross-section and through a commercial orbit-boost mission.
\end{abstract}


\section{Introduction}
\label{sec:intro}

The Neil Gehrels Swift Observatory\cite{gehrels2004} (hereafter Swift) was launched on 20 November 2004 with the primary goal of detecting and rapidly localizing gamma-ray bursts (GRBs) across multiple wavelengths. More than twenty years later, Swift remains one of the most productive and oversubscribed observatories in space, and a cornerstone of the rapidly growing field of time-domain and multi-messenger astrophysics (TDAMM). What is remarkable about Swift's longevity is not simply that the hardware continues to function, but that the mission's \emph{operational} capabilities have grown enormously over its lifetime, allowing it to take on science cases that were never envisioned at launch.

This evolution is no accident. The Swift Science Operations Team (SOT) and Flight Operations Team (FOT), both hosted at The Pennsylvania State University (PSU), operate under an explicit philosophy that the mission's software and procedures are never to be considered fixed. Nearly every day, some part of Swift's operations is changed to make the observatory better at what it does. In this paper we describe how this culture of continuous development has repeatedly reinvented the way Swift is commanded, and how each innovation has unlocked new science. We focus in particular on the path from Swift's original design as an autonomous, GRB-chasing robotic telescope to its present role as a flexible, highly automated platform capable of responding to external triggers, from gravitational-wave detectors and neutrino observatories to fast radio burst monitors, in seconds.

Section~\ref{sec:swift} describes the Swift observatory and what makes both the spacecraft and its operations unique. Section~\ref{sec:evolution} traces the operational evolution of the mission, from the early loss of active X-ray Telescope cooling through the development of automated TOO uploads, onboard tiling, and the ManyPoint capability. Section~\ref{sec:mma} describes how multi-messenger astrophysics has acted as a change driver, and Section~\ref{sec:urgency0} describes the ``Urgency 0'' continuous-commanding mode and its science. Section~\ref{sec:future} discusses the future, including the lessons Swift offers to future missions and the ongoing effort to extend Swift's orbital lifetime.

\section{The Neil Gehrels Swift Observatory}
\label{sec:swift}

\subsection{The Instruments}
\begin{figure}[t]
\centering
\includegraphics[height=8.5cm]{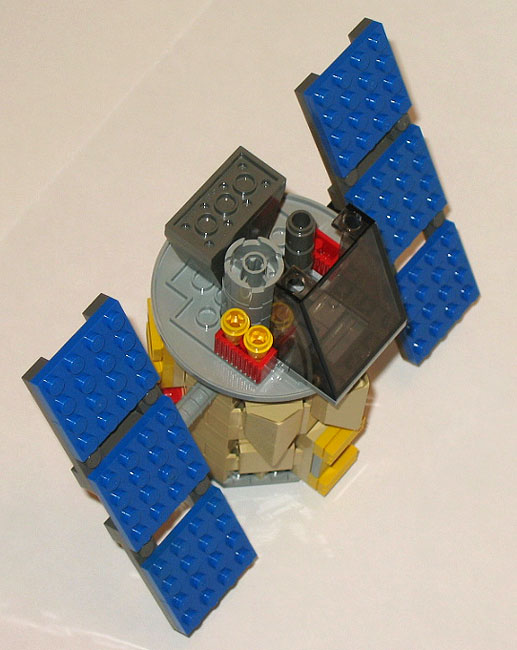}
\caption[Swift instruments]{\label{fig:instrument} A scale model of the Neil Gehrels Swift Observatory, illustrating the compact, co-aligned arrangement of its three instruments: the Burst Alert Telescope (BAT, 15--150~keV), the X-Ray Telescope (XRT, 0.3--10~keV), and the UV/Optical Telescope (UVOT, 170--650~nm). The simultaneous hard X-ray, soft X-ray, and UV/optical coverage of a single target that this arrangement provides remains a space-unique capability.}
\end{figure}

Swift carries three co-aligned instruments that together provide simultaneous hard X-ray, soft X-ray, and ultraviolet/optical coverage of a target, a capability that remains unique in space (Fig.~\ref{fig:instrument}). The Burst Alert Telescope\cite{barthelmy2005} (BAT) is a coded-aperture ``hard X-ray'' instrument operating in the 15--150~keV band, using a large array of CdZnTe detectors to view 2~sr of the sky (roughly one sixth of the celestial sphere) at any time. At launch BAT triggered on approximately 100 GRBs per year (as the instrument aged, this number went down) and computes positions to an accuracy of $\sim$1--3 arcminutes. The X-Ray Telescope\cite{burrows2005} (XRT) is a focusing ``soft X-ray'' instrument covering 0.3--10~keV with a 23.8 arcminute diameter field of view, capable of CCD spectroscopy and of localizing sources to a few arcseconds (as good as 1.8\arcsec with astrometric corrections). The UV/Optical Telescope\cite{roming2005} (UVOT) observes from 170--650~nm over a 17 arcminute square field of view, providing sub-arcsecond positions, six broad-band filters, grism spectroscopy, and a sensitivity of $\sim$22nd magnitude.

\subsection{Detecting a Gamma-Ray Burst}
The canonical Swift observing sequence demonstrates the power of this instrument package. When BAT triggers on a GRB and calculates a position to 1--3 arcminutes, it sends a notification to the team and the spacecraft autonomously slews to point the narrow-field instruments at the burst within 1--2 minutes. The XRT obtains a prompt position of $\sim$5--6\arcsec, refined to $\sim$1.3--3.5\arcsec\ after a few minutes, while UVOT images the field. The Swift team analyzes the data in real time and distributes a Circular through NASA's General Coordinates Network (GCN; \cite{GCN}) to the community within roughly 5--20 minutes.

\subsection{What Makes Swift Unique}
Several features distinguish Swift from other facilities. Its \emph{multi-wavelength} design provides space-unique simultaneous hard X-ray, soft X-ray, and UV coverage. Its capacity for \emph{transient discovery} is anchored by BAT, which acts as a hard X-ray all-sky monitor (covering the full sky in a day), triggering on GRBs, soft gamma repeaters (SGRs), low-mass X-ray binaries (LMXBs), and supergiant fast X-ray transients (SFXTs), while the BAT transient monitor tracks the brightness of hundreds of X-ray sources and discovers new ones. \emph{Rapid slewing} not only gets Swift to a GRB quickly but enables an exceptionally high operational efficiency ($\sim$72\%) and high-cadence, high-sensitivity monitoring for time-domain astrophysics. Finally, Swift operates an exceptionally \emph{open} Target of Opportunity (TOO) program, with very low rejection rates, and makes its data public as soon as possible.

\subsection{What Makes Swift Operations Unique}
Equally important is how Swift is operated. Uniquely among NASA astrophysics missions, all of Swift's operations are hosted at a single university. The FOT, run by Omitron under contract to PSU, consists of aerospace engineers responsible for spacecraft commanding and health and safety. The SOT, run by Penn State Astronomy \& Astrophysics faculty and staff, aided by Swift's international partners in the UK and Italy, schedules observations and TOOs, handles GRBs, implements the Guest Investigator program (which is managed by NASA's Goddard Space Flight Center) by scheduling its targets and TOOs, and interfaces with the community. The co-location of the FOT and SOT fosters a uniquely collaborative environment in which the FOT feels genuine ownership of the mission's scientific success, and is therefore willing to attempt ambitious new modes of operation to meet science needs. The operational configuration is never regarded as fixed, and improvements are made on a near-daily basis.

\subsection{A Growing Demand for Target-of-Opportunity Observations}
\begin{figure}[t]
\centering
\includegraphics[width=0.78\textwidth]{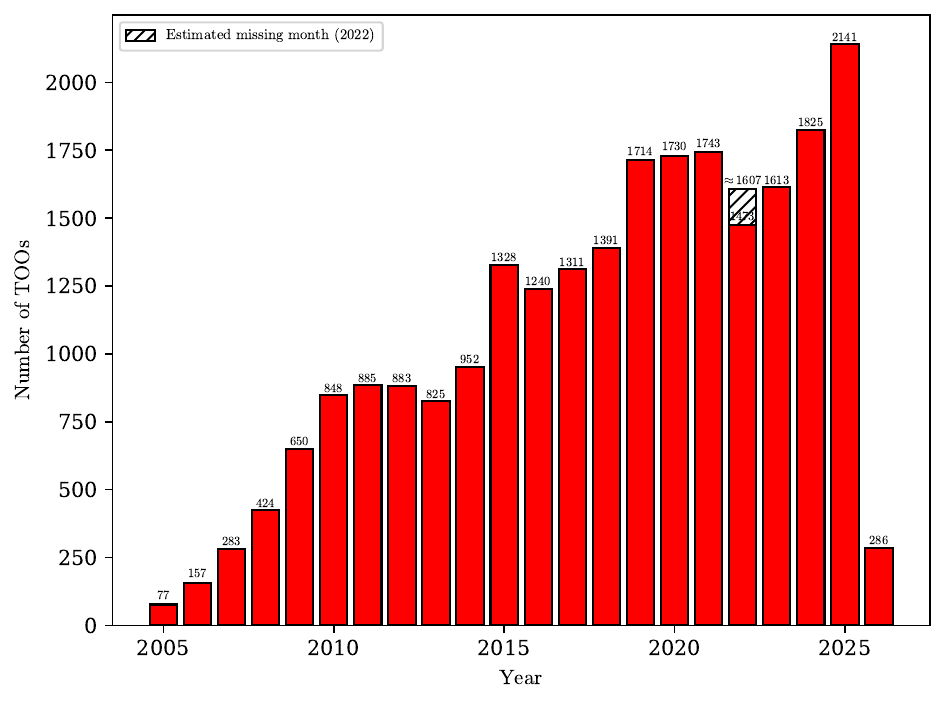}
\caption[TOO requests per year]{\label{fig:toos} The number of TOO requests
submitted to Swift per year. The high numbers reflect the openness of the program, its
$\sim$99\% acceptance rate, the growth of automated submissions through the
Swift TOO API, and in 2025 the recent arrival of new partner missions such as EP and
SVOM. Swift science operations were suspending in February 2026 for a planned
orbit-boost operation, which accounts for the drop in requests that year.}
\end{figure}

The openness of Swift's TOO program has produced a steadily rising demand
(Fig.~\ref{fig:toos}). In 2024 Swift received 1{,}825 TOO requests in a single
year, and in 2025 the total was 2{,}141, a jump likely
driven in part by the recent launches of the Einstein Probe (EP; \cite{EP}) and
the Space-based multi-band astronomical Variable Objects Monitor (SVOM;
\cite{SVOM}). This demand reflects several factors: the program is open to
anyone in the world; the acceptance rate has been $\sim$99\% for the entire
mission; an increasing fraction of requests arrive automatically through the
Swift TOO Application Programming Interface (API); and Swift remains hugely
capable for a broad cross-section of the astronomical community.

A notable fraction of these requests, roughly one third, originate from optical surveys: the discovery of supernovae, tidal disruption events, and novae has historically driven about one third of all TOO requests to Swift. As the Vera C. Rubin Observatory begins its Legacy Survey of Space and Time, the rate of optical transient discovery is expected to surge, and Swift is well positioned to perform space-unique X-ray and UV follow-up of the transients that Rubin uncovers.

Submitting a TOO is deliberately straightforward. Users may employ the web-based TOO page at \url{https://www.swift.psu.edu/toop}, which is itself just a client for the underlying API, or they may use the Python \texttt{swifttools} package and the Swift TOO API directly (\url{https://www.swift.psu.edu/too_api}), allowing fully programmatic automated requests.

\section{The Operational Evolution of Swift}
\label{sec:evolution}

\subsection{An Autonomous Robotic Telescope in Space}
At its core, Swift is an autonomous robotic telescope. It responds automatically to onboard triggers from BAT, slewing to observe GRBs and other transient sources such as SGRs, LMXBs, and high-mass X-ray binaries. While waiting for triggers, Swift observes targets from a Pre-Planned Science Timeline (PPST), generated by the SOT and uploaded daily on weekdays, typically covering one or two days as needed to span weekends. The Swift Mission Operations Center (MOC) is staffed only from 8\,am to 5\,pm, Monday through Friday; however, ground-station passes occur around the clock, so from the very beginning the MOC was automated to allow passes to execute without anyone present.

Swift can be commanded to observe a TOO target using only two spacecraft commands: a first command that sets the exposure time of the TOO response, and a second that sets the coordinates, the instrument modes (a single one- or two-byte number per instrument), and a figure of merit. Only targets in the PPST with an equal or lower figure of merit are displaced by a TOO or GRB; GRBs are assigned a merit of 100 by default. This compact, merit-based commanding scheme proved to be a powerful foundation upon which later automation was built.

\subsection{A Forcing Function: The Loss of Active XRT Cooling}
Perhaps the most consequential early event in Swift's operational history was a hardware failure. The XRT was designed to operate at $-100$\,$^{\circ}$C at all times, achieved through a combination of passive radiative cooling and an active thermo-electric cooler (TEC). Early in the mission, before the XRT had even been cooled, the TEC power supply failed. Without the TEC, the XRT operated between $-40$\,$^{\circ}$C and $-75$\,$^{\circ}$C. Above $-50$\,$^{\circ}$C the XRT dark current became too high and its data unusable, so the team had to learn how to keep the detector cold using passive means alone.

The team discovered that the XRT temperature was driven primarily by the exposure of the XRT radiator to the warm Earth. By orienting the spacecraft so that the radiator faced away from the Earth, the XRT could be kept below $-50$\,$^{\circ}$C through careful construction of the PPST. The problem was that when a GRB went off, the spacecraft would slew autonomously to an arbitrary orientation, and ``all bets were off'' as far as thermal control was concerned. Managing XRT temperature in the face of unpredictable GRB slews became a persistent operational challenge, and, as described below, a key driver of automation.

\subsection{TOO Automation: First Steps}
Driven by the need to manage XRT overheating following GRBs, and by the rising popularity of TOOs, the FOT developed a method to automate TOO uploads. Initially, this automation applied only to ground-station passes: the SOT would generate a ``TOO file'' and submit it to the MOC computer over a secure link, and the MOC computer would then uplink the commands at the next ground-station pass, all without a human in the loop. This was an important first step toward a fully automated NASA transient mission, because it meant TOOs could be executed by the SOT around the clock without needing to call the FOT into the MOC outside of working hours. It also gave the team a tool to manage thermal risk: when a GRB occurs, it is automatically assessed for its ``hotness,'' i.e., whether it will heat the XRT excessively, and if so the Observatory Duty Scientist (ODS) can upload a command at the next ground-station pass to lower its merit, adjust the roll angle, or cancel the observation entirely.

\subsection{Onboard Tiling of Large Error Regions}
\begin{figure}[t]
\centering
\includegraphics[height=6.5cm]{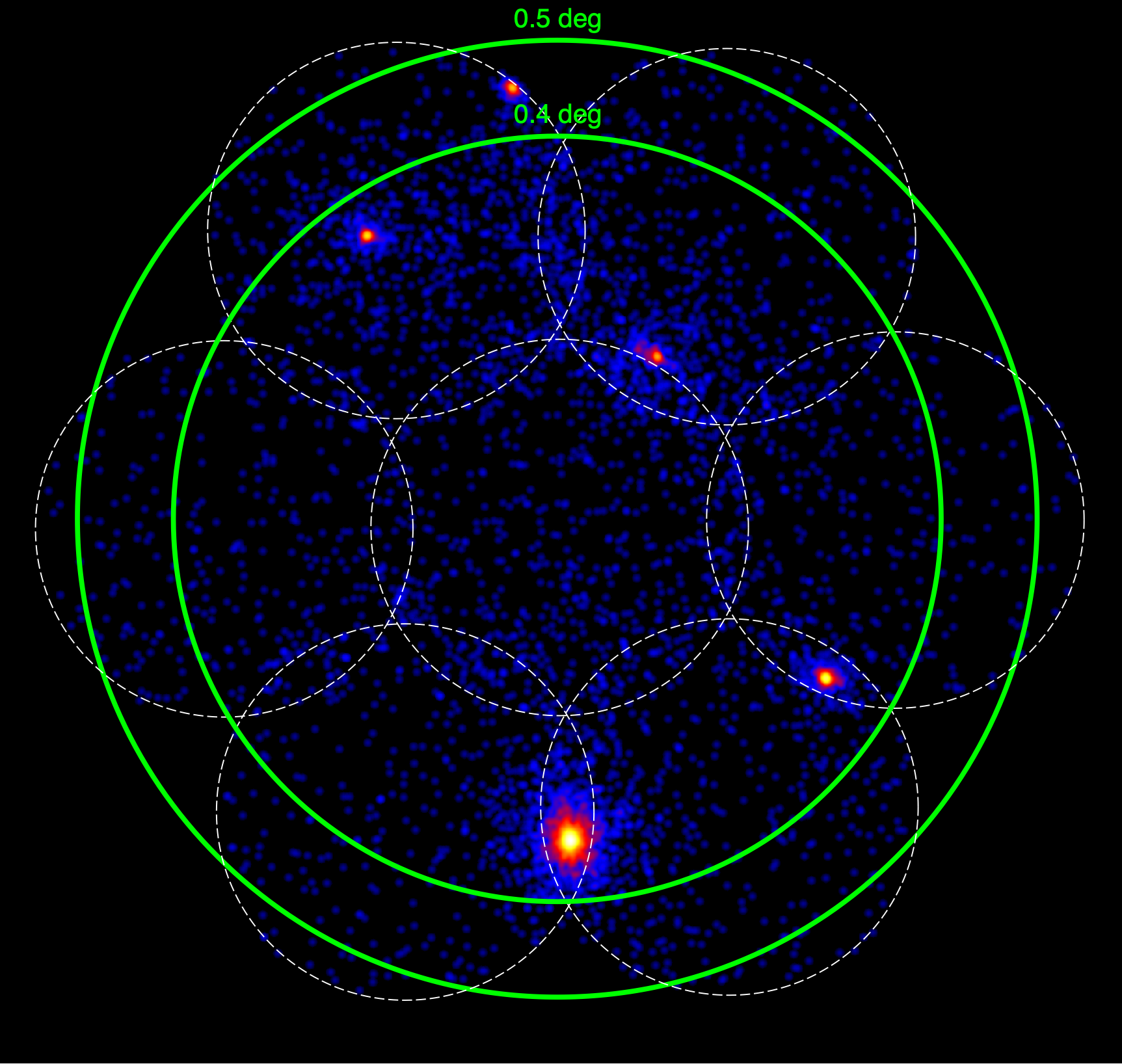}
\caption[Onboard tiling]{\label{fig:tiling} Swift's onboard tiling patterns, used to cover error regions larger than the XRT field of view. Predefined 4-, 7-, 19-, and 37-point patterns cover roughly circular regions up to $\sim$1\,degree across; recent enhancements allow custom, non-circular tiling shapes for optimal coverage of irregular error regions.}
\end{figure}

Although many transients are well localized and fit within the XRT field of view, some are not, and require coverage of a larger area. To address this, the Swift team developed an onboard tiling capability that can cover larger error regions using a predefined pattern of pointings (Fig.~\ref{fig:tiling}). The program includes 4-, 7-, 19-, and 37-point tilings arranged in a roughly circular pattern, capable of covering error regions up to about one degree across. More recent operational innovations have introduced custom tiling shapes, so that non-circular error regions, such as the complex localizations produced by gravitational-wave detectors, can now be tiled optimally.

\subsection{ManyPoint: Executing Complex Pointing Sequences}
To meet the need for ever more complex TOOs, including gravitational-wave tilings involving hundreds of pointings, the Swift team developed a new flight-software capability called ``ManyPoint.'' Rather than uploading a single TOO command, ManyPoint allows the team to upload a list of commands together with start and stop times. For safety, the commands are restricted to only TOO and TOO~ABORT, so ManyPoint cannot be used for other spacecraft commanding, and ManyPoint files can only be uplinked during ground-station passes. In a typical pass, a ManyPoint file containing 100--200 commands can be uplinked.

ManyPoint introduced several key innovations. It allows observations to be scheduled to begin at a chosen later time, which is valuable for coordinating with other facilities. It allows strict control of the start and stop times of TOOs, which makes them both safer, by avoiding problematic slews, and thermally cooler. And it allows a single upload to command observations of up to hundreds of distinct targets. ManyPoint became an essential enabling technology for Swift's gravitational-wave follow-up campaigns.

\section{A Change Driver: Multi-Messenger Astrophysics}
\label{sec:mma}

\subsection{Swift's Long History in Multi-Messenger Astrophysics}
Swift's principal investigator, Neil Gehrels, recognized early the scientific promise of gravitational-wave (GW) astronomy, and as a result Swift has been engaged in multi-messenger astrophysics longer than most facilities. The first Swift observations of LIGO triggers were carried out in 2010, when the team followed up the ``Big Dog'' and ``January'' blind-injection events, well before the Advanced LIGO era\cite{evans2012}. Swift's first follow-up of neutrino triggers began in 2011, with observations of IceCube events\cite{evans2015}. These early campaigns established both the scientific case and the operational tools that would prove essential when real detections began.

\subsection{GW150914: Pushing the Limits of Early Software}
The first direct GW detection, GW150914, was a binary black-hole merger. Unfortunately for Swift, BAT was looking almost in the opposite direction from the source, and the northern lobe of the localization fell within Swift's Sun constraint. Swift nonetheless performed a hexagonal 37-point tiling that, by post-facto calculation, covered approximately 0.3\% of the probability region\cite{abbott2016}. This was the limit of what Swift's software could handle at the time and made clear that, for future GW triggers, the team would need to substantially improve its tiling capability, motivating the development of the custom tiling and ManyPoint capabilities described above.

\subsection{GW170817 and AT2017gfo: A Triumph of Tiling}
\begin{figure}[t]
\centering
\includegraphics[width=0.95\textwidth]{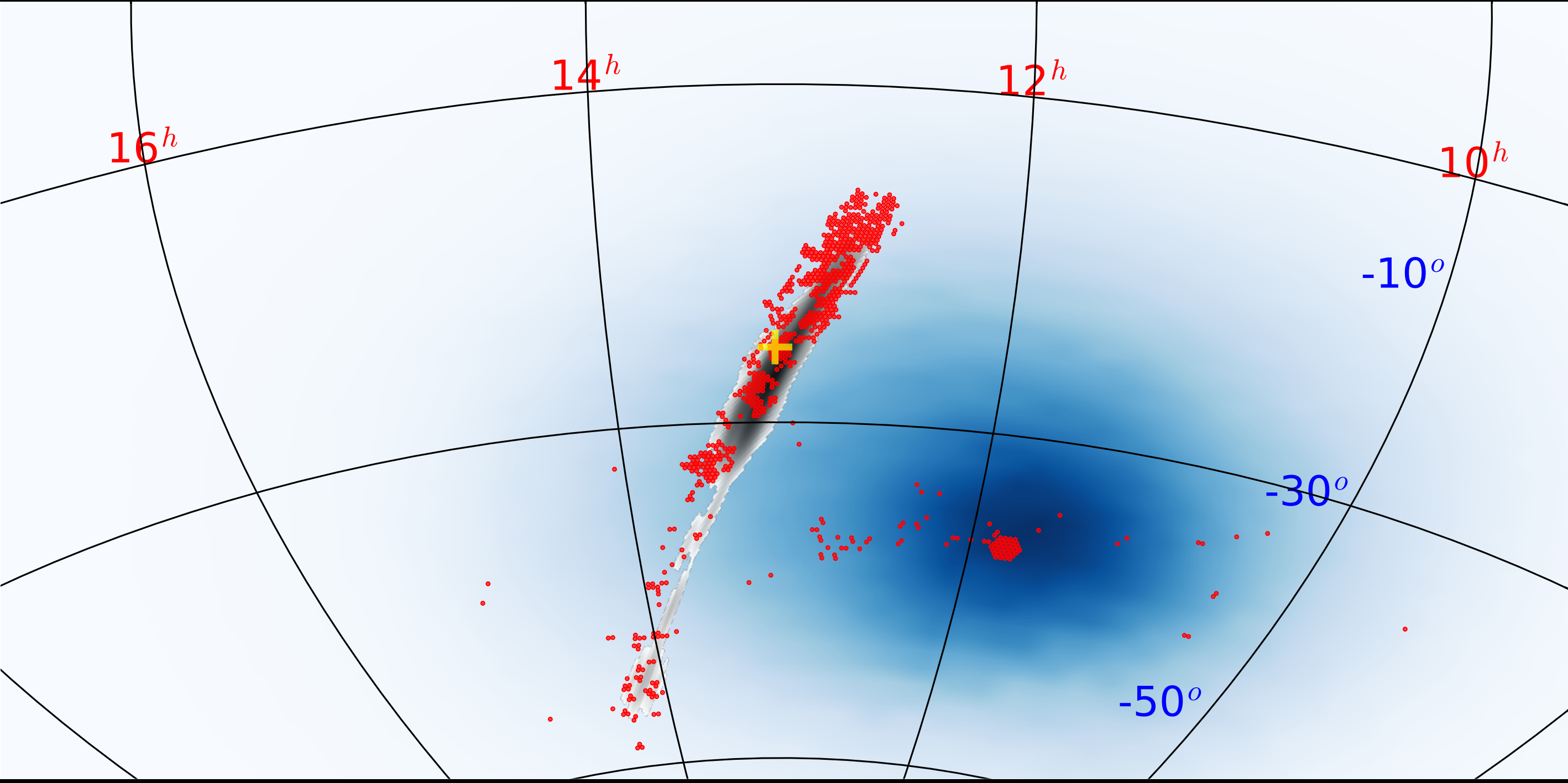}
\caption[GW170817 tiling]{\label{fig:gw170817} Swift's tiled follow-up of the binary neutron-star merger GW170817. Swift observed 744 fields, covering 92\% of the distance-weighted gravitational-wave localization. Galaxy-targeted tiling of this scale by a spacecraft is a uniquely Swift capability\cite{evans2017}.}
\end{figure}

The binary neutron-star merger GW170817 was a landmark for multi-messenger astronomy, and Swift's response demonstrated the maturity of its tiling capability. Swift observed 744 fields, covering 92\% of the distance-weighted GW localization (Fig.~\ref{fig:gw170817}). This style of galaxy-targeted tiling by a spacecraft remains a uniquely Swift capability\cite{evans2017}.

\begin{figure}[t]
\centering
\includegraphics[width=0.95\textwidth]{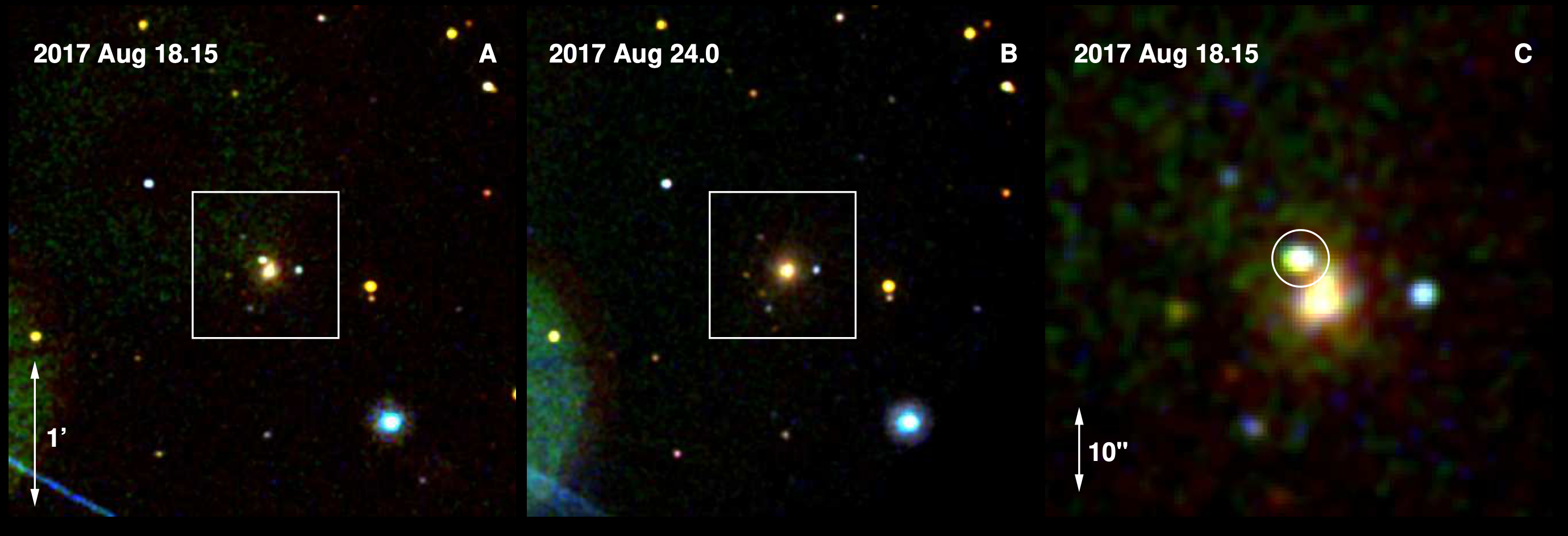}
\caption[AT2017gfo]{\label{fig:at2017gfo} Swift/UVOT detection of the kilonova AT2017gfo, the optical/UV counterpart to GW170817, in NGC~4993. The initial measurement was $u = 18.2 \pm 0.1$~mag (AB). Monitoring showed the ultraviolet emission fading rapidly, while no X-ray emission from the optical transient was detected by the XRT in the early epochs\cite{evans2017}.}
\end{figure}

Swift's UVOT detected the optical/UV counterpart, AT2017gfo, in the galaxy NGC~4993 (Fig.~\ref{fig:at2017gfo}), with an initial measurement of $u = 18.2 \pm 0.1$~mag (AB). Monitoring showed that ultraviolet emission faded rapidly, a key observational signature of the kilonova, while the XRT did not detect X-ray emission from the optical transient by the early epochs\cite{evans2017}.

In the LIGO/Virgo O3 run, Swift followed up the binary black-hole merger S200224A, covering 79.2\% of the GW error region in X-rays and 62.4\% in the UV. No counterpart candidates were seen, and the observations placed an upper limit on the isotropic-equivalent blast-wave energy of $4.1 \times 10^{51}$~erg, assuming GRB-like parameters\cite{klinger2020}.

\subsection{IceCube-170922A: Localizing a Neutrino Source}
Swift's neutrino follow-up reached a milestone with the extremely-high-energy (EHE) event IceCube-170922A, detected at 20:54:30.43~UT on 22 September 2017. A Swift automated TOO was generated directly from the EHE alert, and a 19-point tiling was performed to cover the error circle, with Swift on target 3.25 hours after the IceCube detection. The blazar TXS~0506+056 was detected by the XRT and reported as a possible counterpart\cite{keivani2018}, an association later supported by Fermi monitoring. Multiple IceCube follow-up programs remain in progress, including the follow-up of well-localized Astrotrack Gold alerts and of Fermi blazars near IceCube triggers.

\subsection{BAT GUANO: Recovering Sub-threshold Bursts}
BAT does not trigger onboard on every GRB within its field of view. The Gamma-ray Urgent Archiver for Novel Opportunities\cite{guano2020} (GUANO) is a system that allows BAT event data to be dumped for sensitive ground-based analysis. BAT continuously records data in event mode, but normally transmits only reconstructed data to the ground; the raw event data are retained on board for only about 30 minutes. Using these event data together with a maximum-likelihood analysis technique, NITRATES\cite{nitrates2022}, the team can recover arcminute localizations for roughly ten Fermi GBM GRBs per year, and detect many more.

Because the event data are live for only 30 minutes, recovering them required an entirely new way of getting a command to the spacecraft to save the data to its solid-state recorder. Using the Tracking and Data Relay Satellite System (TDRSS) and API access, the team developed a method to request a TDRS forward link within 14 minutes of a trigger, without a human in the loop, a capability that itself foreshadowed the very low latency commanding described in the next section.

\section{Very Rapid Response: The ``Urgency 0'' Era}
\label{sec:urgency0}

\subsection{Chasing Fast Radio Bursts}
The drive toward ever-faster response is well illustrated by Swift's fast radio burst (FRB) follow-up. Swift triggers on alerts from CHIME and other FRB detectors; if the error region is smaller than the XRT field of view ($\sim$12\arcmin\ radius), the team attempts to observe with the XRT and UVOT as rapidly as possible using the ``AutoTOO'' system. An early demonstration achieved follow-up at $T_0 + 32$ minutes, and two additional FRBs were observed at $T_0 + 39$ minutes and $T_0 + 21$ minutes. In 2023, advances in Swift operations reduced this latency to $T_0 + 3.9$ minutes, and in 2025, after further refinement of the AutoTOO system, Swift was on target for a CHIME FRB just $T_0 + 82$ seconds after the trigger (although in that instance a CHIME glitch had delivered the wrong coordinates). This progression, from tens of minutes to under two minutes, set the stage for the continuous-commanding capability that followed.

\subsection{Urgency 0: Continuous Commanding}
In 2023, in preparation for the LIGO/Virgo/KAGRA (LVK) O4 run, Swift enabled ``Continuous Commanding,'' an operational mode known as ``Urgency 0,'' which reduced TOO latencies from hours (the previous typical best being 14 minutes) to seconds. This opened up entirely new science cases and dramatically increased Swift's responsiveness to TDAMM science. The Urgency 0 capability makes it possible to attempt ``early-warning'' slewing to increase the odds of a coincident BAT detection of a neutron-star merger\cite{tohuvavohu2024}; it enables GW tiling to commence much more quickly; and it allows rapid follow-up of counterparts identified by other telescopes. Critically, it places triggers from other missions on a level playing field with triggers from BAT itself. The ability to immediately command a TDAMM spacecraft to observe has many use cases and greatly increases scientific return, and in demonstrating it Swift is creating vital heritage for future mission proposals.

\subsection{Urgency 0 GRB Science with Partner Missions}
Swift now performs as-soon-as-possible observations of GCN alerts from the SVOM ECLAIRs instrument and the EP Wide-field X-ray Telescope (WXT). Swift's XRT and UVOT enable rapid identification of the source class and arcsecond-resolution localization. Results are published over GCN and made available live at dedicated websites for EP (\url{https://www.swift.ac.uk/EP}) and SVOM (\url{https://www.swift.ac.uk/SVOM}). In doing so, Swift has proven that it can perform rapid GRB observations of triggers not only from its own BAT but also of triggers from other missions. This points to a future in which advanced, continuous communications make it unnecessary to physically couple a GRB detector with its follow-up telescopes on the same spacecraft.

\subsection{Urgency 0 Statistics}
\begin{figure}[t]
\centering
\includegraphics[width=0.7\textwidth]{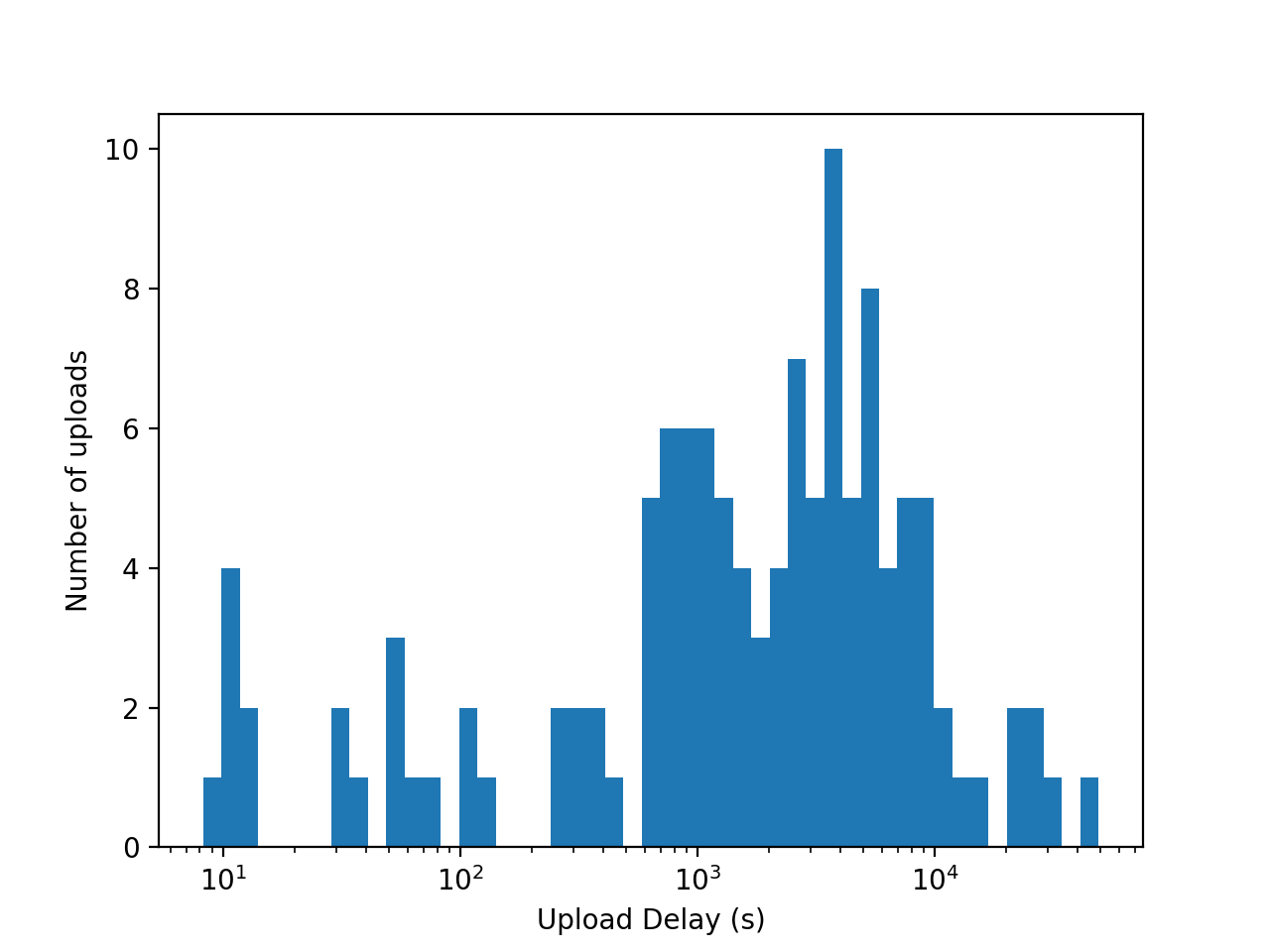}
\caption[Urgency 0 response times]{\label{fig:urgency0} Distribution of Swift ``Urgency 0'' continuous-commanding response times across the 155 automated uploads performed to date. The fastest responses are $\sim$10~s from receipt of the TOO at the MOC to the start of slew; the median of $\sim$2.2~ks reflects the fact that only 25--42\% of the sky is observable at any instant due to Earth, Moon, and Sun constraints.}
\end{figure}

To date, Swift has performed 155 Urgency~0 automated uploads (Fig.~\ref{fig:urgency0}). The fastest response times have been of order $\sim$10~seconds, measured from the receipt of the TOO at the MOC to the start of the slew. The median response time is $\sim$2.2~ks, reflecting the fact that Swift can see only 25--42\% of the sky at any given moment, with 58\% blocked by Earth and the remainder by Moon and Sun constraints.

\subsection{Chasing Gravitational Waves in Real Time}
Tohuvavohu et al. \cite{tohuvavohu2024} proposed a novel use of the Urgency~0 capability: to attempt to catch a GW event with BAT \emph{as the merger occurs}, using ``early-warning'' alerts from LVK. Achieving this requires reducing latency as much as possible on the LVK side for early-warning alerts; being able to respond extremely rapidly at the Swift MOC; calculating the optimum way to align BAT with the LVK error region; and performing a GUANO dump at the time of the merger.

Working with the LVK group, the team developed and tested a low-latency pipeline to channel early-warning GW events through to the Swift system, along with software to process these events and uplink commands in real time, with processing times of 0.2--1.7~s and an uplink time of $\sim$10~s. The system has already triggered. The first TOO command was received at 19:12:39.524~UT; the first TOO command was written to the database $\sim$289~ms later at 19:12:39.813; the command was sent to the MOC computer and received by the spacecraft about 10~s after that (the bit rate makes this $\sim$10~s essentially always); and the slew alert was received through the spacecraft's ITOS telemetry $\sim$13 [ The predicted merger time, $T_0$, was 19:13:00.641, so the spacecraft was commanded to slew at $T_0$ minus 7~s. Although this particular trigger turned out not to be real, it provided a powerful end-to-end demonstration of the process. The next generation of GW detectors will be able to detect neutron-star mergers up to 15 minutes before merger, making it essential to demonstrate this capability now.

\section{The Future}
\label{sec:future}

\subsection{Swift Will Not Last Forever}
Swift and Fermi were rightly identified by the Astro2020 Decadal Survey\cite{astro2020} as essential for TDAMM in the future. Yet both observatories are old, and Swift's orbit is decaying: absent intervention, the spacecraft is likely to re-enter the atmosphere in late 2026. The techniques and methods that Swift has introduced over the last 20 years have therefore created vital heritage that will enable new missions to take on its role in the future. However, for this to happen, it is essential that the lessons learned from Swift are not forgotten.

\subsection{Key Lessons for Future Missions}
A central lesson of Swift's operational evolution is that the next ``Swift'' need not be a single observatory; it could instead be a collaboration of many missions, provided they share certain key capabilities. First, constant communications mean that the triggering instrument and the follow-up instrument no longer need to reside on the same spacecraft, as Swift's Urgency~0 follow-up of EP and SVOM has demonstrated. Second, dedicated GRB-detection missions are valuable: an instrument that is not constantly slewing to perform TOOs can achieve a higher duty cycle and is better suited to detecting ultra-long and faint, high-redshift GRBs; future GRB detectors are likely to be lower-cost, stand-alone instruments such as the BlackCAT CubeSat\cite{Chattopadhyay18}. Third, rapid X-ray follow-up to obtain arcsecond afterglow positions remains essential, and could be provided by dedicated smaller missions whose primary purpose is rapid response, free of the competing science pressures that burden larger general-purpose X-ray observatories. Fourth, rapid UV/optical follow-up may be served from space by UVEX\cite{Kulkarni21} and ULTRASAT\cite{Shvartzvald24}, with optical and infrared coverage well provided from the ground and additional rapid-response space telescopes planned. Finally, and crucially, Swift's experience shows the value of an open TOO policy and an open data policy: these ensure a broad audience and a greater number of scientific papers. Having multiple groups publish on the same event is a feature, not a bug, and open collaboration between missions, as Swift has practiced most recently with EP and SVOM, is key to maximizing science return and to ensuring a mission's continued vitality.

\subsection{The Swift Boost Mission}
Although there is a real need to develop the replacement for Swift, Swift may not yet be done. NASA is actively working to restore Swift to a higher orbit ($\geq$550~km). This is necessary because, after more than twenty years on orbit, Swift's orbit has decayed to the point that re-entry is expected in late 2026. NASA has selected Katalyst Space Technologies to perform the boost; %(Fig.~\ref{fig:katalyst})
with a planned launch timeframe of mid-2026, an extraordinarily fast turnaround for a mission of this kind. A successful boost would extend the operational life of a still-unique and highly productive observatory at a critical moment for time-domain and multi-messenger astrophysics.

\subsection{Reducing Swift's Atmospheric Drag Cross-Section}
\label{sec:drag}
The same philosophy that has reshaped Swift's transient response has more recently been turned toward the problem of orbital decay itself. The motivation became urgent in February 2026, when an updated prediction of Swift's orbital decay indicated that the spacecraft could re-enter the atmosphere \emph{before} the planned boost mission (Section~\ref{sec:future}) would be in position to rendezvous with it. Keeping Swift on orbit long enough for that rendezvous to take place therefore became a priority, and, with no propulsion of its own, the only lever available to the team was the spacecraft's attitude. Because Swift has no propulsion, the rate at which its orbit decays is set by atmospheric drag, which is in turn proportional to the cross-sectional area the spacecraft presents to its direction of travel (the ram direction). That area is not fixed: it depends strongly on the spacecraft's attitude, because Swift's large solar-array wings sweep out very different projected areas depending on how the observatory is oriented relative to its velocity vector. By preferentially choosing science attitudes that present a smaller cross-section to the ram direction, the time-averaged drag can be reduced, slowing the decay of the orbit and buying valuable additional operational lifetime, entirely through changes in operations rather than hardware.
 
To exploit this, the SOT has developed a drag-reduction scheduler that folds drag minimization directly into the daily generation of the PPST. For each observing slot, the scheduler evaluates the candidate pointings that satisfy Swift's observing constraints and selects those that minimize a combined cost metric. The drag cross-section itself can be estimated with models of increasing fidelity, ranging from a simple geometric box approximation, through an analytic $L_p$-norm model, to a high-accuracy look-up table derived from detailed spacecraft geometry. Crucially, drag is not optimized in isolation: the scheduler balances it against the other operational realities that have always governed Swift planning. A slew-distance penalty discourages large, inefficient (and thermally costly) slews; an Earth-Elevation-Angle term biases the schedule toward colder attitudes to protect instrument temperatures, reusing the very thermal-management insight that the loss of active XRT cooling forced upon the team two decades earlier (Section~\ref{sec:evolution}); and solar-panel illumination is tracked to ensure that power generation is not compromised. The resulting optimization metric is, schematically, the sum of the predicted drag area and these attitude penalties, so that the chosen timeline is the one that is simultaneously low-drag, thermally safe, power-positive, and observationally efficient.
 
The team has also built diagnostics to quantify the benefit, computing the time-averaged drag cross-section of any given timeline so that competing schedules can be compared directly and the realized improvement measured. This drag-aware scheduling complements the more dramatic intervention described below: where a reboost would restore altitude in a single step, attitude-based drag reduction continuously slows the loss of altitude in the interim. In the near term its specific purpose is to keep Swift on orbit long enough for the boost mission to reach and rendezvous with it. It is a clear demonstration that, for Swift, operational innovation extends not only the mission's scientific reach but its very lifetime.

\subsection{Extending Swift's Life: The Boost Mission}
In parallel with the drag-reduction effort, NASA is leading an active mission to reboost Swift to a higher orbit, with a requirement of 550~km and a goal of 600~km. This is necessary because the strong solar maximum drove Swift's orbit to decay faster than expected, to the point that, without intervention, re-entry would occur in the near term. NASA has selected Katalyst Space Technologies to perform the boost using a servicing spacecraft named LINK. The concept of operations calls for LINK to launch, complete system checkouts, phase and rendezvous with Swift, perform proximity operations and capture, execute the orbit-raise burn, and then release Swift to resume operations, with the servicer extending Swift's mission life by an estimated ten years. Remarkably, the timeframe from Katalyst's selection to launch was less than a year, an extraordinarily fast turnaround for a mission of this kind, and the NASA and Katalyst teams behind it deserve enormous credit. A successful boost would extend the operational life of a still-unique and highly productive observatory at a critical moment for time-domain and multi-messenger astrophysics.

\subsection{Swift Boost: Current Status}
As of this writing, the boost mission is well under way. Katalyst's LINK servicer launched on a Pegasus~XL rocket on 3~July 2026. The launch was successful and LINK is on orbit, power-positive, and communicating with the ground, with commissioning expected to continue for a few weeks. Rendezvous with Swift is expected in August~2026, with the boost phase itself anticipated to last roughly three months, followed by recommissioning and a return to full science operations. If successful, this will make Swift one of the first science missions to have its life extended by a commercial servicing spacecraft, a milestone whose operational lessons will themselves become part of Swift's heritage for the missions that follow.

\section{Conclusions}
\label{sec:conclusions}

For more than two decades, the Neil Gehrels Swift Observatory has remained at the forefront of time-domain and multi-messenger astrophysics, not because its hardware is new, but because its operations are never allowed to stand still. Through a sustained culture of continuous development, enabled by the unique co-location of flight and science operations at a single university, Swift has transformed itself from an autonomous GRB-chasing telescope into a flexible, highly automated platform that can respond to triggers from across the electromagnetic and multi-messenger landscape in seconds. The early loss of active XRT cooling forced the development of automated TOO uploads; the demands of gravitational-wave follow-up drove onboard tiling and the ManyPoint capability; and the needs of the LVK O4 era produced the Urgency~0 continuous-commanding mode, which has enabled real-time, early-warning slewing toward predicted mergers. These innovations have delivered landmark science, from the UVOT detection of the kilonova AT2017gfo to the localization of the neutrino blazar TXS~0506+056, and the same culture is now being applied to the mission's own survival: attitude-based drag-minimizing operations have measurably slowed Swift's orbital decay and kept it aloft long enough for a commercial servicer, Katalyst's LINK, which launched successfully on 3~July 2026, to attempt the first reboost of a science observatory in low Earth orbit. Whether or not that boost ultimately succeeds, the operational heritage Swift has created, together with its demonstrated model of openness and inter-mission collaboration, offers an essential template for the future of rapid-response astrophysics.

% References
\acknowledgments
The author thanks the entire Swift Science Operations Team and Flight Operations Team, past and present, whose continuous innovation over more than twenty years is the subject of this paper, and acknowledges the foundational vision of the late Neil Gehrels. The Swift mission is operated by The Pennsylvania State University under contract to NASA's Goddard Space Flight Center.

\end{document}